
\input harvmac


\Title{\vbox{\baselineskip12pt\hbox{EFI-92-66}}}
{\vbox{\centerline{Light-Cone Quantization of }
	\vskip2pt\centerline{the Liouville Model }}}

\centerline{Jadwiga Bie\'{n}kowska\footnote{$^\dagger$}
{jadwiga@yukawa.uchicago.edu}}
\bigskip\centerline{Department of Physics and}
\centerline{Enrico Fermi Institute}
\centerline{The University of Chicago}\centerline{5640 S. Ellis Ave Chicago ,
IL 60637}


\vskip .3in
We present the quantization of  the Liouville model
defined in light-cone coordinates in (1,1) signature
space. We   take advantage of the
 representation of the Liouville field by the free field of the Backl\"{u}nd
 transformation and adapt the approch by Braaten, Curtright and Thorn
{\ref\bct{E. Braaten, T. Curtright and C. Thorn, Ann. of Phys. 147 (1983) 365
\semi
 E. Braaten, T. Curtright, G. Ghandour and C. Thorn, Ann. of Phys. 153 (1984)
147.}} .
 Quantum operators of the Liouville field $\partial_{+}\phi$,
$\partial_{-}\phi$, $e^{g\phi}$, $e^{2g\phi}$ are constructed consistently in
 terms of the free field. The Liouville model
field theory space is found to be restricted to the
sector  with  field momentum $P_{+}=-P_{-}$, $P_{+}> 0$ , which is
a closed subspace for the Liouville theory operator algebra.
\Date{11/92} 

 \baselineskip=20pt plus 2pt minus 2pt


\newsec{Introduction}
The Liouville model has attracted attention recently for two main reasons.
First the Liouville theory shows up in covariant quantization of the
relativistic string
{\ref\pol{A. M. Polyakov, Phys. Lett. B (1981)  207.}} .
 Second, it describes
a two dimensional  gravity {\ref\dkd{F. David, Mod. Phys. Lett. A3 (1988) 1651
\semi J. Distler and H. Kawai, Nucl. Phys. B321 (1989).}}
 {\ref\seiberg{N. Seiberg, Prog. Theor. Phys.
 Supp. 102 (1990) 319.}}.
The classical Liouville  action

\eqn\llagr
{S_{L}=\int d^{2}x({1\over2}\phi\partial^{\mu}\partial_{\mu}\phi -{(2m)^2\over
{2g^{2}}}e^{2g\phi})
}
with solutions given  by the Liouville equation
\eqn\lequ
{\partial_{-}\partial_{+}\phi= {(2m)^2 \over g}e^{2g\phi},
}
where $\phi$ is a Liouville field,
describes the gravity theory of
 two-dimensional surfaces with constant curvature.

The classical theory \llagr\ is conformally invariant and the improved
stress-energy
tensor is found to have the form
 {\ref\hoker{E. D'Hoker and R. Jackiw
Phys. Rev D 26 (1982) 3517.}}
{\bct}

\eqna\stresslfcl
$$\eqalignno
{T_{++}&=(\partial_{+}\phi)^{2}-{1\over g}\partial_{+}^{2}\phi &\stresslfcl
a\cr
T_{--}&=(\partial_{-}\phi)^{2}-{1 \over g}\partial_{-}^{2}\phi .& \stresslfcl
b\cr
}$$

We expect that the construction of a quantum Liouville theory will help to
understand  quantum gravity at least in the case of simple
two dimensional models {\dkd}.
Liouville type theories also show up  in the  CGHS model {\ref\cghs{ C. G.
Callan, S. B. Giddings, J. A. Harvey and A. Strominger,
Phys, Rev. D45 (1992) R1005.}}
{\ref\sus{J. G. Russo, L. Susskind, L. Thorlacius, preprints
 SU-ITP-92-4, SU-ITP-92-17 \semi L.Susskind, L. Thorlacius,
 preprint SU-ITP-92-12.}}
 {\ref\alwis{ S. P. de Alwis,
preprints COLO-HEP-280. May 1992, COLO-HEP-284,  June 1992.
COLO-HEP-288 July 1992.}}
  {\ref\bilal{ A. Bilal, C. G. Callan, preprint PUPT-1320, May 1992}} which
tries to
address the black hole evaporation  problem .

E. Braaten, T.  Curtright and C. Thorn {\bct} have constructed
the quantum  Liouville theory defined on  Minkowski-type space
with a spatial dimension compactified  on a circle using
 the equal time quantization
prescription. There have also been different approaches to quantize the
Liouville model and some of them have developed in  unexpected ways
{\ref\ger{ J. L. Gervais and A. Neveu, Nucl. Phys B199 (1982) 59. Nucl. Phys.
B209 (1982) 125.
Nucl. Phys. B224 (1983) 329. Nucl. Phys. B238 (1984) 125, 396. Phys. Lett. 151
B (1985) 271.}}
{\ref\jack{E. D'Hoker, D. Z. Freedman and R. Jackiw, Phys. Rev D 28 (1983)
2583.}} leading to the development of  quantum groups theory.

The consistency of quantizing the Liouville theory on the affine Minkowski
space in the
light-cone coordinates  was proven in {\ref\mans{
P.  Mansfield, Nucl. Phys B208 (1982), 277. Nucl. Phys. B222 (1983) 419  \semi
H. C. Liao and P.Mansfield, Nucl. Phys. B344 (1990), 696.}}
. However there has been no
attempt to construct the quantum field theory   operators for the affine
Minkowski space Liouville model.
The use of the light-cone quantization simplifies the analysis of the
massless field theory and we expect that it will also help to
simplify  description of the  Liouville model through
 Backl\"und transformation.

In this paper we quantize the Liouville theory in the light-cone coordinates
and construct quantum operators using the regularized Backl\"und
transformation.
We require that quantum operators of the theory
transform consistently under the conformal  algebra of the stress-energy
tensor.
We  check the consistency  of the quantum equations of motion
described in terms of quantum operators defined by the regularization of
their classical analogs.
The physical states space of the Liouville theory is restricted by the
equations of motion to be
only half the space of the free pseudoscalar field theory and is
described by the vacuum states eigenvalues $P_{+}=-P_{-}$ and $P_{+} >  0$.
 This is in agreement
with the result  of
Braaten, Curthright and Thorn \bct\
 obtained for the Liouville model defined on a circle.

\newsec{Classical Backl\"{u}nd Transformation}

In  light-cone coordinates the Backl\"und transformation for the
 Liouville equation reads:
\eqna\bakeq
$$\eqalignno{
&\partial_{+} \phi= \partial_{+} \psi - {2m \over g} e^{g\phi}e^{g\psi} &
\bakeq a\cr
&\partial_{-} \phi= -\partial_{-} \psi - {2m \over g} e^{g\phi}e^{-g\psi} . &
\bakeq b
}$$
where $\psi(x^{+},x^{-})= \psi(x^{+})+\psi(x^{-})$ is a free field.
The functions $\psi(x^{+})$ and $\psi(x^{-})$ are
completely independent.

To obtain the integral representation of the Liouville field in terms of the
free field  $\psi$ we first integrate the equation \bakeq{a}\  along the
$x^{-}=\,$const light cone from $ x^{+}_{0}=-\infty$ to $x^{+}$ and then
integrate
the  equation \bakeq{b}\ on the  light cone defined by
$x^{+}=x_{0}^{+}=\,$const from $x^{-}_{0}=-\infty$ to $x^{-}$.
Combining these two integrated equations together we
get the integral representation for the classical Liouville field :

\eqn\bakf
{\eqalign{
e^{-g \phi(x^{+},x^{-})}&=2m(\int_{x_{0}^{+}}^{x^{+}}dy^{+}e^{2g\psi(y^{+})}
e^{g\psi(x^{-})-g\psi(x^{+})} +
\int_{x_{0}^{-}}^{y^{-}}dy^{-}e^{-2g\psi(y^{-})}
e^{g\psi(x^{-})-g\psi(x^{+})} \cr
&+e^{-g\phi(x^{+}_{0},x^{-}_{0})}e^{-g\psi(x^{-}_{0})+g\psi(x^{+}_{0})}
e^{g\psi(x^{-})-g\psi(x^{+})} ).
}}

To completely specify the solutions to the Backl\"und equations \bakeq\  we
must impose the boundary conditions for the Liouville field  at  $(x^{+}_{0},
x^{-}_{0})$. The boundary condition  most appropriate for this
 problem is the one where the Liouville potential approaches zero
at spatial and time negative infinity i.e. $e^{2g\phi(x^{+}_{0},x^{-}_{0})}
\rightarrow 0$.
This condition is equivalent to the statement that the Liouville field
approaches the free scalar field in this limit and  is  given by
$\phi(x^{+}_{0},x^{-}_{0}) \rightarrow -\psi(x^{-}_{0})+\psi(x^{+}_{0})$.
With this boundary condition the classical representation of the Backl\"und
transformation is

\eqn\bakcl
{\eqalign{e^{-g
\phi(x^{+},x^{-})}&=2m(\int_{x_{0}^{+}}^{x^{+}}dy^{+}e^{2g\psi(y^{+})}
e^{g\psi(x^{-})-g\psi(x^{+})} +
\int_{x_{0}^{-}}^{x^{-}}dy^{-}e^{-2g\psi(y^{-})}
e^{g\psi(x^{-})-g\psi(x^{+})} \cr
&+e^{g\psi(x^{-})-g\psi(x^{+})} ) .
}}

\newsec{Light-Cone Quantization of the Liouville Model.}

We can quantize the Liouville $\phi$ and free $\psi$ fields on the light cone
plane defined by
$x^{-}$=const using the Dirac prescription for quantization of
systems with dynamical constraints
 {\ref\dirac{ P. A. M. Dirac, ``Lectures on Quantum Mechanics'' (1964). Belfer
Graduate School of Science , Yeshiva University, New York \semi
 A. Hanson, T. Regge and C. Teitelboim,
``Constrained Hamiltonian Systems'', lectures (1976), Academia Nazionale dei
Lincei, Roma.}}
 {\ref\leut{H. Leutwyler and J.  Stern, Ann. of Phys. 112 (1978) 94.}}
. Following the standard procedure  we get the
commutation relations for the fields at fixed $x^{-}$ :

\eqn\fcom{
[\phi(x^{+}),\phi(y^{+})]=[\psi(x^{+}),\psi(y^{+})]=-i{1 \over 4}
\epsilon(x^{+}-y^{+})
}
where $\epsilon(x)={\rm sign}(x)$.

The Fourier representations of the free field in terms of the annihilation and
creation
operators is
\eqn\fourierff
{\psi(x^{+})=\int_{0}^{\infty} {dp \over \sqrt{4\pi} \, p} ( a(p)e^{-ipx^{+}}+
a^{\dagger}(p)e^{ipx^{+}})
}
where the annihilation and creation operators obey the commutation relations
\break
$[a(p),a^{\dagger}(q)]=p\delta(p-q)$. For quantization at the $x^{+}$=const
 plane, the
  $\psi(x^{-})$ part  of the free field obeys the analogous commutation
relation and has a similar Fourier expansion in terms of the $b(p),\,\,
b^{\dagger}(p)$
operators which commute with   $a(p),\,\, a^{\dagger}(p)$. The $\phi$ field can
be
 represented in the same manner
\eqn\fourierlf
{\phi(x^{+})=\int_{0}^{\infty} {dp \over \sqrt{4\pi} \, p} ( d(p)e^{-ipx^{+}}+
d^{\dagger}(p)e^{ipx^{+}})
}

In the representation
{\fourierlf } of the Liouville field the quantum stress energy
tensor is defined as the normal ordered version of the classical \stresslfcl\
stress-energy
tensor
\eqn\stresslfq
{T^{\phi}_{++}(x^{+})=:(\partial_{+} \phi(x^{+}))^{2}:
-{1\over\gamma}\partial^{2}_{+}\phi(x^{+})
}
where the normal ordering is defined with respect to the $d(p)$ and
 $ d^{\dagger}(p)$
annihilation and  creation operators.
It is easy to check that  the stress-energy tensor \stresslfq\
obeys the conformal algebra commutation relations
\eqn\confalg
{[T(x^{+}),T(y^{+})]=(T(x^{+})+T(y^{+}))\,i\,\delta '(x^{+}-y^{+})- C\,i\,
\delta ''' (x^{+}-y^{+})
}
 where C is quantum improved conformal anomaly $C={1 \over 2 \gamma^{2}} + {1
\over 16\pi}$.
 ${1 \over \gamma}$ is the quantum improved conformal factor
 given by the renormalized coupling constant
  ${ \gamma}= g\,(1+{g^{2} \over 2\pi})^{-1}$. The
renormalization of the  Liouville model coupling constant $g$
is easily derived from general considerations   (see
{\ref\dhoker{ E. D'Hoker  ``Lecture Notes on 2D Quantum Gravity and Liouville
Theory'' , proceedings of VI-th Swieca Summer School, Brazil, January 17-28,
1991. }} ).
This is in  agreement with the results found before
in  \bct\   \hoker\ \mans\  .

\newsec{ Quantum Backl\"und Transformation}

In order to determine how the quantum version of the classical Backl\"und
transformation looks
we will adapt the approach developed in {\bct}.
The quantum operators corresponding to the classical ones defined in equation
\bakcl\ are
determined by the normal ordering prescription.

Using the representation of the Liouville field in terms of the creation and
annihilation
 operators \fourierlf\ we can define the normal ordering of an exponential of
the
Liouville field as
\eqn\nord{
N_{\phi}(e^{\alpha\phi(x^{+})})=e^{\alpha\phi^{+}(x^{+})}e^{\alpha\phi^{-}(x^{+})}
}
 where
\eqn\cranlf
{\eqalign{
\phi^{-}(x^{+})&=\int_{0}^{\infty} {dp \over \sqrt{4\pi} \, p}
d(p)e^{-ipx^{+}} \cr
\phi^{+}(x^{+})&=\int_{0}^{\infty} {dp \over \sqrt{4\pi}\,  p}
d^{\dagger}(p)e^{ipx^{+}}.
}}

It is easy to check {\ref\brych{Yu. A. Brychkov and A. P. Prudnikov
``Integral Transforms of Generalized Functions'', ed. Gordon and Breach Science
Publishers S.A., Amsterdam (1989).}}
 that the fields defined above  obey the commutation relation

\eqn\comcran{
[\phi^{-}(x^{+}),\phi^{+}(y^{+})]=-{1 \over 4\pi}{\rm ln}|x^{+}-y^{+}|.
}
Any  conformal field $\Phi_{\Delta_{+}}$ with the left dimension $\Delta_{+}$
obeys the
following commutation relation with the left stress energy tensor

\eqn\dimeql
{i[T_{++}(x^{+}),\Phi_{\Delta_{+}}(y^{+})]=\delta(x^{+}-y^{+})\partial_{y^{+}}\Phi(y^{+})
-\Delta_{+}\delta '(x^{+}-y^{+})\Phi_{\Delta_{+}}.
}

The normal ordered with respect to $d(p),\,\,d^{\dagger}(p)$
$N_{\phi}(e^{-g\phi})$
operator has a  dimension
determined by its commutation relation with quantum stress energy tensor
\stresslfq\ .
Using the formula \dimeql\ it is straightforward to find that the dimension of
the
operator $N_{\phi}(e^{-\alpha\phi(x^{+})})$ is equal to
 $\Delta_{+}={-\alpha \over 2\gamma}
-{\alpha^{2} \over 8\pi}$. From the consistent construction of the quantum
Backl\"und transformation
we  require  that the same operator expressed in terms of the free field
$\psi(x_{+})$ would
have the same conformal dimension with respect to the $\psi$ field
stress-energy tensor.

We postulate then that the quantum version of the Backl\"und transformation
has the form

\eqn\bakquf
{\eqalign{N_{\phi}e^{-g\phi(x^{+},x^{-})}&=2m\,N_{\psi}[e^{g\psi(x^{-})-g\psi(x^{+})}(1 \cr
&+\int_{-\infty}^{x^{+}}dy^{+}f(x^{+}-y^{+})
e^{2g\psi(y^{+})} +\int_{-\infty}^{x^{-}}dy^{-}e^{-2g\psi(y^{-})})]
}}
and the right hand side of the equation \bakquf\ is normal ordered with
 respect to  $a(p),\,\,a^{\dagger}(p)$
 annihilation  and creation operators.

The stress energy tensor for the $\psi$ field is
\eqn\stressffq
{T^{\psi}_{++}(x^{+})=:(\partial_{+} \psi(x^{+}))^{2}:
-\beta_{+}\partial^{2}_{+}\psi(x^{+})
}
The normal ordering prescription for exponentials is defined as
 for the Liouville field \nord\ and the
creation and annihilation parts of the free field $\psi(x^{+})$ have the
commutation relations
\eqn\ancrcom
{[\psi^{-}(x^{+}),\psi^{+}(y^{+})]=-{1 \over 4\pi}{\rm ln}|x^{+}-y^{+}|.
}

 The requirement that the quantum
Liouville field exponential defined by the transformation \bakquf\ in the
$\psi$
representation is the
conformal field of the same dimension as in the $\phi$ field representation
fixes
$f(x^{+}-y^{+})$ and $\beta_{+}$ conformal improvement factor for free field.
It is  straightforward to check that
$f(x^{+}-y^{+})=(x^{+}-y^{+})^{{g^{2} \over 2\pi}}$ and
   $\beta_{+}={1 \over \gamma}$, which
is the same as the conformal improvement factor for the quantum
Liouville field stress energy tensor \stresslfq\ .

The same line of argument can be followed for the $x^{-}$ coordinate
independently.
Even if we do not know the commutation relations of the Liouville field
 $\phi(x^{+},x^{-})$ itself for arbitrary points in the space-time
 (neither $x^{+}$ nor $x^{-}$  held fixed),
  we know that
the free fields $\psi(x^{+})$ and $\psi(x^{-})$ commute with each other and so
we can construct the complete expression for the quantum
$e^{-g\phi}$ operator in terms of the free field $\psi$.
Repeating the same calculation for the right handed part of the operator
$N_{\phi}(e^{-g\phi})$
we obtain that
 $f(x^{-}-y^{-})=(x^{-}-y^{-})^{{g^{2} \over 2\pi}}$
 and $\beta _{-}=-{1 \over \gamma}$.
This  confirms the results obtained before \bct\ that field $\psi$ has to be a
pseudo-scalar
 free field.

The derivation of the quantum version of the Backl\"und transformation
can be summarized  by the formula

\eqn\bakqu
{\eqalign{N(e^{-g\phi(x^{+},x^{-})})&=2mN_{\psi}[e^{g\psi(x^{-})-g\psi(x^{+})}(1 \cr
&+\int_{-\infty}^{x^{+}}dy^{+}(x^{+}-y^{+})^{{g^{2} \over 2\pi}}
e^{2g\psi(y^{+})} +\int_{-\infty}^{x^{-}}dy^{-}(x^{-}-y^{-})^{{g^{2} \over
2\pi}}e^{-2g\psi(y^{-})})]
.}}

To completely define the set of quantum operators in this  language
we have to specify
 the normal
ordering prescription for two exponential operators
 of the Liouville field. The natural definition is:

\eqn\exnord{
N(e^{\alpha\phi}e^{\beta\phi}) ={\rm lim}_{x \rightarrow y}
|x^{+}-y^{+}|^{{\alpha \beta} \over {4\pi}}
|x^{-}-y^{-}|^{{\alpha \beta} \over {4\pi}}N(e^{\alpha\phi(x^{+},x^{-})})
N(e^{\beta\phi(y^{+},y^{-})}).
}

Using the equation \exnord\ with $\alpha=g$, $\beta=-g$ we find the quantum
operator

\eqn\exfg
{\eqalign{
Ne^{ g\phi(x^{+},x^{-})}&={1\over 2m}e^{g\psi^{+}(x^{+})}e^{-g\psi^{+}(x^{-})}
X(x^{+},x^{-})e^{g\psi^{-}(x^{+})}e^{-g\psi^{-}(x^{-})}
}}

where

\eqn\exfgx
{\eqalign{
X^{-1}(x^{+},x^{-})=&N_{\psi}[1+\int_{-\infty}^{x^{+}}dy^{+}
(x^{+}-y^{+})^{-{g^{2} \over 2\pi}}
e^{2g\psi(y^{+})} \cr
&+\int_{-\infty}^{x^{-}}dy^{-}(x^{-}-y^{-})^{-{g^{2} \over
2\pi}}e^{-2g\psi(y^{-})}]
.}}

 The same way we find that

\eqn\exsg
{\eqalign{Ne^{2g\phi(x^{+})}&={1\over
(2m)^{2}}e^{2g\psi^{+}(x^{+})}e^{-2g\psi^{+}(x^{-})}
\tilde{X}^{\dagger}(x^{+},x^{-})\tilde{X}(x^{+},x^{-})
e^{2g\psi^{-}(x^{+})}e^{-2g\psi^{-}(x^{-})}
}}

where

\eqn\exsgx
{\eqalign
{\tilde{X}^{-1}(x^{+},x^{-})=&N_{\psi}[1+\int_{-\infty}^{x^{+}}dy^{+}(x^{+}-y^{+})^{-{g^{2} \over \pi}}
e^{2g\psi(y^{+})} \cr
&+\int_{-\infty}^{x^{-}}dy^{-}(x^{-}-y^{-})^{-{g^{2} \over
\pi}}e^{-2g\psi(y^{-})}]
.}}

 From  definitions \exfgx\ and \exsgx\ it is clear that these formulas do not
contain the integration singularities as long as ${g^{2} \over \pi} < 1$ . They
are valid in the
 weak coupling limit of the Liouville model. This restriction is identical to
the one obtained
for the Liouville model with  space dimension compactified on a circle {\bct}.

The complete set of  operators includes also $\partial_{+}\phi$ and
$\partial_{-}\phi$
 which are defined by the classical Backl\"und transformations \bakcl\ .
Looking at the equations \bakcl\ and \exfg\
 we may guess  the form of the quantum version of the
classical operators $e^{g\phi}e^{g\psi}$ and  $e^{g\phi}e^{-g\psi}$

\eqna\exglgf
$$\eqalignno
{N(e^{g\phi}e^{g\psi})&={1\over
2m}e^{2g\psi^{+}(x^{+})}Y_{+}(x^{+},x^{-})e^{2g\psi^{-}(X^{+})}&
\exglgf a \cr
N(e^{g\phi}e^{-g\psi})&={1\over 2m}e^{-2g\psi^{+}(x^{-})}Y_{-}(x^{+},x^{-})
e^{-2g\psi^{-}(x^{-})} . &\exglgf b
}$$

Requiring that the equation of motion \lequ\  hold also for the quantum
operators
we find that $Y_{+},\,\,Y_{-}$ are given by the expressions

\eqna\exglgfy
$$\eqalignno
{Y^{-1}_{+}(x^{+},x^{-})&=N_{\psi}[1+\int_{-\infty}^{x^{+}}dy^{+}(x^{+}-y^{+})^{-{g^{2} \over \pi}}
e^{2g\psi(y^{+})}
+\int_{-\infty}^{x^{-}}dy^{-}e^{-2g\psi(y^{-})}]\cr
&  &\exglgfy a \cr
Y^{-1}_{-}(x^{+},x^{-})&=N_{\psi}[1+\int_{-\infty}^{x^{+}}dy^{+}
e^{2g\psi(y^{+})}
+\int_{-\infty}^{x^{-}}dy^{-}(x^{-}-y^{-})^{-{g^{2} \over
\pi}}e^{-2g\psi(y^{-})}] . \cr
& &\exglgfy b
}$$

We can also check, using the equation \dimeql\ and its right-handed
counterpart,
  that the left and right conformal dimensions
of the operators
 $N(e^{g\phi}e^{g\psi})$ and $N(e^{g\phi}e^{-g\psi})$ are respectively
$\Delta_{+}=1$
and $\Delta_{-}=1$ which proves the consistency of the definitions \exglgf\  .

The quantum equations of motion are then described by
\eqn\eqmq{\eqalign{
&\partial_{+}\partial_{-}\phi=\partial_{-}(\partial_{+}\psi
 -{2m \over g}N(e^{g\phi}e^{g\psi})) \cr
&\partial_{+}(-\partial_{-}\psi -{2m \over g}N(e^{g\phi}e^{-g\psi}))=
{(2m)^{2} \over g}N(e^{2g\phi})
}}
and expressions \exglgf\ , \exsg\ .

\newsec{Liouville Equation  on the Physical States Space}

It was proven in \bct\ that the physical states  space for the Liouville model
defined on a circle is half the space of the free field theory.
 The constraints
of the  Liouville theory space came from the requirement that quantum
equations of motion are valid. It turns out that they hold only on half of
the free field theory space. In this section we would like to establish if
similar restrictions apply to our quantization procedure.

In order to do so
we first  have to define the Fock space of the theory. As usual the Fock space
is defined by the states built up from the creation operators acting on the
unique vacuum state which is annihilated by all the annihilation operators
$a(p)|0\rangle = b(p)|0\rangle =0$.
Special attention however has to be paid to the constant (coordinate
independent)
modes of the
free field operator which as in the case of quantization on the time-like
plane   define different super-selection sectors of the vacuum.
The constant modes of the free field operator are implicitly contained
in the Fourier expansion \fourierff\ . We can recover  constant modes of the
free field operator as
\eqn\qff{\eqalign{
Q^{+}={\rm lim}_{p\rightarrow 0}{1\over {p\sqrt{4\pi}}}(a(p)+a^{\dagger}(p))
\cr
Q^{-}={\rm lim}_{p\rightarrow 0}{1\over {p\sqrt{4\pi}}}(b(p)+b^{\dagger}(p)) .
\cr
}}
These operators have their canonical conjugates  defined by
\eqn\pff{\eqalign{
P_{+}={\rm lim}_{p\rightarrow 0}{i \sqrt{\pi}}(-a(p)+a^{\dagger}(p)) \cr
P_{-}={\rm lim}_{p\rightarrow 0}{i \sqrt{\pi}}(-b(p)+b^{\dagger}(p)) . \cr
}}
They obey the canonical commutation relations
\eqn\cancomm{
[Q^{+},P_{+}]=i {\hskip 2cm} [Q^{-},P_{-}]=i
}

With constant mode operators and their canonical conjugates given by the
formulas
\qff\ and \pff\ we may define normalized vacuum states of the theory as

\eqn\vac{\eqalign{
e^{iQ^{+}P'_{+}}e^{iQ^{-}P'_{-}}|0\rangle&=|P'_{+},P'_{-},0\rangle \cr
 {\rm and} {\hskip 1cm}
\langle 0 P'_{+},P'_{-}|P''_{+},P''_{-},0\rangle &=
\delta(P'_{+}-P''_{+})\delta(P'_{-}-P''_{-})
}}
where zero in the $|P'_{+},P'_{-},0\rangle$ means a state annihilated by all
$a(p),\,\,b(p)$ with $p\neq 0$.

We can also rewrite the Fourier expansion of the free field as
\eqn\ffqp{\eqalign{
\psi(x^{+}) &=Q^{+}+P_{+}{x^{+}\over 2\pi}+\int_{0^{+}}^{\infty} {dp \over
\sqrt{4\pi} p}
 ( a(p)e^{-ipx^{+}}+
a^{\dagger}(p)e^{ipx^{+}})= \cr
 &= Q^{+}+P_{+}{x^{+}\over 2\pi}+\tilde{\psi}(x^{+}) \cr
\psi(x^{-}) &=Q^{-}+P_{-}{x^{-}\over 2\pi}+\int_{0^{+}}^{\infty} {dp \over
\sqrt{4\pi} p}
 ( b(p)e^{-ipx^{-}}+
b^{\dagger}(p)e^{ipx^{-}})= \cr
 &= Q^{-}+P_{-}{x^{-}\over 2\pi}+\tilde{\psi}(x^{-}) . \cr
}}

 New fields $\tilde{\psi}(x^{+}),\,\,\tilde{\psi}(x^{-})$ contain only nonzero
momentum creation, annihilation modes and have slightly modified commutation
relations

\eqn\crancom{\eqalign{
[\tilde{\psi}^{-}(x^{+}),\tilde{\psi}^{+}(y^{+})]&=-{1 \over 4\pi}{\rm
ln}|x^{+}-y^{+}|+
{i\over 4\pi}(x^{+}-y^{+}) \cr
[\tilde{\psi}^{-}(x^{-}),\tilde{\psi}^{+}(y^{-})]&=-{1 \over 4\pi}{\rm
ln}|x^{-}-y^{-}|+
{i\over 4\pi}(x^{-}-y^{-}).
}}

We have to specify the normal ordering prescription for the new set
 of the operators
including the ``momenta" and ``position" operators on the field space. For the
nonzero momentum operators the normal ordering prescription is as usual. For
the
$Q^{+},\,\,P_{+},\,\,Q^{-},\,\,P_{-}$ operators we define the normal ordering
by

\eqn\norord{\eqalign{
N(e^{2\alpha Q^{+}}F(P_{+}))=e^{\alpha Q^{+}}F(P_{+})e^{\alpha Q^{+}} {\hskip
1cm}
N(e^{2\alpha Q^{-}}F(P_{-}))=e^{\alpha Q^{-}}F(P_{-})e^{\alpha Q^{-}} .
}}
 It is easy to check that this normal ordering prescription differs only
by a coordinate independent
constant from the one given in equation \nord\ .

The Liouville model operators defined by the modified ordering prescription
have the form
\eqn\exfgqp
{\eqalign{
Ne^{ g\phi(x^{+},x^{-})}&={1\over
2m}e^{g\tilde{\psi}^{+}(x^{+})}e^{-g\tilde{\psi}^{+}(x^{-})}
X_{Q}(x^{+},x^{-})e^{g\tilde{\psi}^{-}(x^{+})}e^{-g\tilde{\psi}^{-}(x^{-})} \cr
X^{-1}_{Q}(x^{+},x^{-})&=N[e^{-g\psi_{0}(x^{+})}e^{g\psi_{0}(x^{-})}(1+
\int_{-\infty}^{x^{+}}dy^{+}(x^{+}-y^{+})^{-{g^{2} \over 2\pi}}
e^{2g\psi(y^{+})} \cr
&+\int_{-\infty}^{x^{-}}dy^{-}(x^{-}-y^{-})^{-{g^{2} \over
2\pi}}e^{-2g\psi(y^{-})})]
}}

and

\eqn\exsgqp
{\eqalign{Ne^{2g\phi(x^{+},x^{-})}&={1\over
(2m)^{2}}e^{2g\tilde{\psi}^{+}(x^{+})}
e^{-2g\tilde{\psi}^{+}(x^{-})}
\tilde{X}_{Q}^{\dagger}(x^{+},x^{-})\tilde{X}_{Q}(x^{+},x^{-})
e^{2g\tilde{\psi}^{-}(x^{+})}e^{-2g\tilde{\psi}^{-}(x^{-})} \cr
\tilde{X}^{-1}_{Q}(x^{+},x^{-})&=N[e^{-g\psi_{0}(x^{+})}e^{g\psi_{0}(x^{-})}(1+
\int_{-\infty}^{x^{+}}dy^{+}(x^{+}-y^{+})^{-{g^{2} \over \pi}}e^{{i\over
2\pi}(x^{+}-y^{+})}
e^{2g\psi(y^{+})} \cr
&+\int_{-\infty}^{x^{-}}dy^{-}(x^{-}-y^{-})^{-{g^{2} \over \pi}}e^{{i\over
2\pi}(x^{-}-y^{-})}
e^{-2g\psi(y^{-})})]
.}}

 The operators involved in the expression of  $\partial_{+}\phi$ and
$\partial_{-}\phi$
are given by

 \eqna\exglgfqp
$$\eqalignno
{N(e^{g\phi}e^{g\psi})&={1\over
2m}e^{2g\tilde{\psi}^{+}(x^{+})}Y_{Q+}(x^{+},x^{-})
e^{2g\tilde{\psi}^{-}(x^{+})}
 & \exglgfqp a \cr
N(e^{g\phi}e^{-g\psi})&={1\over
2m}e^{-2g\tilde{\psi}^{+}(x^{-})}Y_{Q-}(x^{+},x^{-})
e^{-2g\tilde{\psi}^{-}(x^{-})}
 &\exglgfqp b
}$$

and

\eqna\exglgfyqp
$$\eqalignno
{Y^{-1}_{Q+}(x^{+},x^{-})&=N[e^{-2g\psi_{0}(x^{+})}(1+
\int_{-\infty}^{x^{+}}dy^{+}(x^{+}-y^{+})^{-{g^{2} \over \pi}}
e^{2g\psi(y^{+})}
+\int_{-\infty}^{x^{-}}dy^{-}e^{-2g\psi(y^{-})})]&\cr
& &\exglgfyqp a \cr
Y^{-1}_{Q-}(x^{+},x^{-})&=N[e^{2g\psi_{0}(x^{-})}(1+\int_{-\infty}^{x^{+}}dy^{+}
e^{2g\psi(y^{+})}
+\int_{-\infty}^{x^{-}}dy^{-}(x^{-}-y^{-})^{-{g^{2} \over
\pi}}e^{-2g\psi(y^{-})})] .& \cr
& &\exglgfyqp b
}$$

In the above expressions we used the shorthand notations for the zero modes of
the free
field
$\psi_{0}(x^{+})=Q^{+}+P_{+}{x^{+} \over 2\pi}$ and
$\psi_{0}(x^{-})=Q^{-}+P_{-}{x^{-} \over 2\pi}$.

To establish the consistency of the quantum equation of motion to
 the lowest order
in the $g$ expansion we have to check if the equation

\eqn\lqu{\eqalign{
\partial_{-}\langle 0,P'_{+},P'_{-}|\partial_{+}\phi|P''_{+},P''_{-},0\rangle&=
\langle 0,P'_{+},P'_{-}|{(2m)^{2} \over g}Ne^{2g\phi}|P''_{+},P''_{-},0\rangle
}}
holds to the lowest order in the $g$ expansion.

 From the equations \exsgqp\ and \exglgfqp\  we see that we need to evaluate
expressions

\eqn\eqmota
{\eqalign{
\partial_{-}\langle 0,P'_{+},P'_{-}|-{1 \over
g}Y_{Q+}(x^{+},x^{-})|P''_{+},P''_{-},0\rangle \cr
{\rm and} {\hskip 1cm}
\langle 0,P'_{+},P'_{-}|{1 \over g}\tilde{X}_{Q}^{\dagger}\tilde{X}_{Q}
|P''_{+},P''_{-},0\rangle .
}}

The evaluation of the equation of motion coming from
the second equation of Backl\"und transformation \bakeq{b}\ can be
 reduced to the
problem of calculating the matrix element
$\langle 0,P'_{+},P'_{-}|-{1 \over
g}Y_{Q-}(x^{+},x^{-})|P''_{+},P''_{-},0\rangle$
which is similar to the \eqmota\  formula
 and we will  concentrate  on examining the consistency of the equation \lqu\
{}.

To the lowest order in $g$ the following formulas hold

\eqn\eqmoty
{\eqalign{
Y_{Q+}^{-1}&=(1+O(g^{2}))(e^{-gQ^{+}}e^{-gP_{+}{x^{+} \over
\pi}}e^{-gQ^{+}}+\cr
&+\int_{-\infty}^{x^{+}}dy^{+}e^{gP_{+}{(y^{+}-x^{+}) \over \pi}}+ \cr
&+e^{-gQ^{+}}e^{-gP_{+}{x^{+} \over \pi}}e^{-gQ^{+}}e^{-gQ^{-}}
\int_{-\infty}^{x^{-}}dy^{-}e^{-gP_{-}{y^{+} \over \pi}}
e^{-gQ^{-}})
}}

\eqn\eqmotx{\eqalign{
\tilde{X}_{Q}^{-1}&=(1+O(g^{2}))(e^{-{g \over 2}Q^{+}}e^{-gP_{+}{x^{+} \over
2\pi}}e^{-{g \over 2}Q^{+}}
e^{{g \over 2}Q^{-}}e^{gP_{-}{x^{-} \over 2\pi}}e^{{g \over 2}Q^{-}}+ \cr
&+e^{{g \over 2}Q^{-}}e^{gP_{-}{x^{-} \over 2\pi}}e^{{g \over 2}Q^{-}}
e^{{g \over 2}Q^{+}}e^{gP_{+}{x^{+} \over 2\pi}}
\int_{-\infty}^{x^{+}}dy^{+}e^{gP_{+}{(y^{+}-x^{+}) \over \pi}}
e^{{i \over 2\pi}(x^{+}-y^{+})}e^{{g \over 2}Q^{+}}+ \cr
&+ e^{-{g \over 2}Q^{+}}e^{-gP_{+}{x^{+} \over 2\pi}}e^{-{g \over 2}Q^{+}}
e^{-{g \over 2}Q^{-}}e^{-gP_{-}{x^{-} \over 2\pi}}
\int_{-\infty}^{x^{-}}dy^{-}e^{gP_{-}{(x^{-}-y^{-}) \over \pi}}
e^{{i \over 2\pi}(x^{-}-y^{-})}e^{-{g \over 2}Q^{-}} ).
}}

The integrations over  $y^{+}$ and $y^{-}$ in the expressions \eqmoty\  ,
\eqmotx\
give a finite result (which is equivalent to
the nonzero matrix elements \eqmota\ ) provided we
are restricted to the space with $P_{+}> 0$ and $P_{-}< 0$ eigenvalues. After
performing the integrations and  using  identities

\eqna\exnorm
$$\eqalignno{
e^{-gQ^{+}}e^{-qP_{+}{x^{+} \over \pi}}e^{-gQ^{+}}&
=e^{iP_{+}^{2}{x^{+} \over 4\pi}}e^{-2gQ^{+}}e^{-iP_{+}^{2}{x^{+} \over 4\pi}}
&\exnorm a\cr
e^{gQ^{+}}P_{+}^{-1}e^{gQ^{+}}&={1 \over 2ig}\Gamma({{P_{+}+ig} \over
2ig})e^{2gQ^{+}}
\Gamma^{-1}({{P_{+}+ig} \over 2ig})& \exnorm b \cr
e^{-gQ^{-}}(-P_{-})^{-1}e^{-gQ^{-}}&={1 \over 2ig}\Gamma^{-1}(-{{P_{-}-ig}
\over 2ig})e^{-2gQ^{-}}
\Gamma(-{{P_{-}-ig} \over 2ig})& \exnorm c \cr
}$$
we get the following expressions for the matrix elements \eqmota\ :

\eqn\exnormy{\eqalign{
\partial_{-}\langle 0,P'_{+},P'_{-}|-{1 \over
g}Y_{Q+}(x^{+},x^{-})|P''_{+},P''_{-},0\rangle&=\cr
={-i \over g}{g \over \pi}{{(P'^{2}_{-}-P''^{2}_{-})} \over 4\pi}
e^{i{(P'^{2}_{-}-P''^{2}_{-})x^{-} \over 4\pi}}
e^{i{(P'^{2}_{+}-P''^{2}_{+})x^{+} \over 4\pi}}P'_{+}\times &\cr
\Gamma({P'_{+} \over 2ig})\Gamma(-{{P'_{-}-ig} \over 2ig})
\Gamma^{-1}({P''_{+} \over 2ig})\Gamma^{-1}(-{{P''_{-}-ig} \over 2ig}) &\times
\cr
 \langle 0,P'_{+},P'_{-}|{e^{2gQ^{+}} \over
{e^{2gQ^{+}}+e^{-2gQ^{-}}}}|P''_{+},P''_{-},0\rangle
 (1+O(g^{2})) &
}}

and

\eqn\exnormx{\eqalign{
\langle 0,P'_{+},P'_{-}|{1 \over g}\tilde{X}_{Q}^{\dagger}\tilde{X}_{Q}
|P''_{+},P''_{-},0\rangle=& \cr
={1 \over g}({g \over \pi})^{2} (2g)^{2}
e^{i{(P'^{2}_{-}-P''^{2}_{-})x^{-} \over
4\pi}}e^{i{(P'^{2}_{+}-P''^{2}_{+})x^{+} \over 4\pi}}
\Gamma^{-1}(-{P'_{+} \over 2ig})\Gamma^{-1}({P'_{-} \over 2ig})
\Gamma^{-1}({P''_{+} \over 2ig})\Gamma^{-1}(-{P''_{-} \over 2ig})\times \cr
\langle 0,P'_{+},P'_{-}|{{e^{gQ^{+}}e^{-gQ^{-}}} \over
{e^{2gQ^{+}}+e^{-2gQ^{-}}}}
\Gamma({P_{+} \over 2ig}+{1 \over 2})\Gamma(-{P_{+} \over 2ig}+{1 \over 2})
\Gamma({P_{-} \over 2ig}+{1 \over 2})\Gamma(-{P_{-} \over 2ig}+{1 \over 2}) &
\cr
{{e^{gQ^{+}}e^{-gQ^{-}}} \over {e^{2gQ^{+}}+e^{-2gQ^{-}}}}
|P''_{+},P''_{-},0\rangle
 (1+O(g^{2})).&
}}

The expressions \exnormy\ and \exnormx\ can be further evaluated quite
strightforwardly.
We insert the complete set of states
$\int {dq^{\pm} \over 2\pi}|q^{\pm}\rangle \langle q^{\pm}|$
or $\int d\tilde{p}^{\pm}|\tilde{p}^{\pm}\rangle \langle \tilde{p}^{\pm}|$ and
after tedious
but standard integrations
we obtain the following expressions:

\eqn\exnormyy{\eqalign{
\partial_{-}\langle 0,P'_{+},P'_{-}|-{1 \over
g}Y_{Q+}(x^{+},x^{-})|P''_{+},P''_{-},0\rangle&=\cr
={1 \over g}{g \over \pi}{{(P'^{2}_{-}-P''^{2}_{-})} \over 4\pi}
e^{i{(P'^{2}_{-}-P''^{2}_{-})x^{-} \over 4\pi}}
e^{i{(P'^{2}_{+}-P''^{2}_{+})x^{+} \over 4\pi}}P'_{+}\times &\cr
\Gamma({P'_{+} \over 2ig})\Gamma(-{{P'_{-}-ig} \over 2ig})
\Gamma^{-1}({P''_{+} \over 2ig})\Gamma^{-1}(-{{P''_{-}-ig} \over 2ig}) &\times
\cr
{1 \over  { 2g \,\, 2{\rm sh}{{\pi\Delta P_{-}}\over 2g}}}\delta(\Delta
P_{+}-\Delta P_{-})
}}

and

\eqn\exnormxx{\eqalign{
\langle 0,P'_{+},P'_{-}|{1 \over g}\tilde{X}_{Q}^{\dagger}\tilde{X}_{Q}
|P''_{+},P''_{-},0\rangle=& \cr
={-1 \over g}({g \over 2})^{2}
e^{i{(P'^{2}_{-}-P''^{2}_{-})x^{-} \over
4\pi}}e^{i{(P'^{2}_{+}-P''^{2}_{+})x^{+} \over 4\pi}}
\Gamma^{-1}(-{P'_{+} \over 2ig})\Gamma^{-1}({P'_{-} \over 2ig})
\Gamma^{-1}({P''_{+} \over 2ig})\Gamma^{-1}(-{P''_{-} \over 2ig})\times \cr
\left({{S_{-}-S_{+}} \over {{\rm sh}{{\pi P''_{+}}\over 2g}\,{
\rm sh}{{\pi P'_{+}}\over 2g}\,{\rm sh}{{\pi(S_{-}-S_{+})}\over 4g}}}
+\right. & \cr
+{{ 2\Delta P_{+}({\rm ch}^{2}{{\pi\Delta P_{+}}\over 4g}\,{\rm sh}{{\pi
S_{+}}\over 4g}\,
{\rm sh}{{\pi S_{-}}\over 4g} +
{\rm sh}^{2}{{\pi \Delta P_{+}}\over 4g}\,{\rm ch}{{\pi S_{+}}\over 4g}\,{\rm
ch}{{\pi S_{-}}\over 4g})}
 \over
{{\rm sh}{{\pi \Delta P_{+}}\over 2g}\,{\rm sh}{{\pi P''_{+}}\over 2g}\,
{\rm sh}{{\pi P'_{+}}\over 2g}\,{\rm sh}{{\pi(-\Delta P_{+} + S_{-})}\over 4g}
\,{\rm sh}{{\pi(\Delta P_{+} + S_{-})}\over 4g}}} + & \cr
\left. -{{S_{-}{\rm sh}{{\pi(S_{-}+S_{+})}\over 4g}} \over
{{\rm sh}{{\pi P'_{+}}\over 2g}\,
{\rm sh}{{\pi P''_{+}}\over 2g}\,{\rm sh}{{\pi(-\Delta P_{+} + S_{-})}\over
4g}\,
{\rm sh}{{\pi(\Delta P_{+} + S_{-})}\over 4g}}}\right) \times
\delta(\Delta P_{+}-\Delta P_{-}).
}}
We have introduced the symbols: $\Delta P_{+}=P''_{+}-P'_{+}$, $\Delta
P_{-}=P''_{-}-P'_{-}$,
 $S_{+}=P''_{+}+P'_{+}$, $S_{-}=P''_{-}+P'_{-}$, sh=sinh and  ch=cosh.

Inspection of  expressions {\exnormyy} and  {\exnormxx} indicates this
 equivalence
 in the small $g$ limit, if the field theory space is restricted to the vacuum
 states $|P'_{+},\,\, P'_{-},\,\,0 \rangle$ with momenta $P'_{+}=-P'_{-}$ ,
 $P'_{+} > 0$.
This is half of the field theory space for the pseudoscalar free field
$\psi$ defined by $P'_{+}=-P'_{-}$ condition.
The leading $g$ contribution to the expression \exnormxx\ comes from the second
term in the sum.

On this  space the matrix elements are equal to{\foot{Useful identities for
$\Gamma$ function
can be found in  {\ref\grad{I. S. Gradshteyn and I. M. Ryzhik,
``Tables of Integrals, Series and Products'', ed. Academic Press, Inc.
 New York (1981).}} }}

\eqn\mx{\eqalign{
\partial_{-}\langle 0,P'_{+},P'_{-}|\partial_{+} \phi
|P''_{+},P''_{-},0\rangle=
\langle 0,P'_{+},P'_{-}|Ne^{2g\phi} |P''_{+},P''_{-},0\rangle= & \cr
={1\over \pi}{ {{P'}_{+}^{2}} \over (2\pi)^{2}} \delta(2\Delta P_{+}) &.
}}

The restriction of the physical states space is consistent with the
integrability
condition for the equations  {\eqmoty} and  {\eqmotx}
 i.e.  $P'_{+} > 0$ and  $P'_{-} < 0$.
The subspace defined above is closed under the operator algebra since acting by
a
Liouville vertex operator with a  positive momentum on the state with positive
momentum will always produce another state with positive momentum.

\newsec{Conclusion}

In this paper we have consistently constructed the quantized Liouville model
defined on the affine Minkowski space.  Aside from
 the fact that this procedure proposes a
 different way of quantizing the theory we   find it interesting because
it can be readily
applied
 to the  CGHS {\cghs} model expressed by Liouville type fields
 {\alwis} {\bilal}.
 The CGHS model is  believed to describe
the properties of evaporating black holes and
quantizing it may provide a new insight into the problem.
Work on this problem is in progress.

\bigskip

I would like to thank Emil Martinec for many helpful discussions and comments,
and
Eric D'Hoker  for discussions about the Liouville theory. This work is
submitted in
partial fulfillment of the requirements for a Ph.D. degree in physics at the
University
of Chicago.
This work is supported in part by the DOE grant DE-FG02-90ER40560.

%
%

\listrefs
\bye